\documentclass[aps,pre,twocolumn,showpacs,nofootinbib,superscriptaddress]{revtex4-1}

\usepackage{amsmath}
\usepackage{amssymb}
\usepackage{graphicx}
  \usepackage{hyperref}
\usepackage[arrow, matrix, curve]{xy}
\usepackage{bm}
\usepackage{yhmath}
\usepackage{color}
\usepackage{url}

\usepackage[usenames,dvipsnames]{xcolor}
\hypersetup{colorlinks=true, linkcolor=BrickRed, urlcolor=blue!50!black, citecolor=blue!50!black}

\usepackage[normalem]{ulem}

\makeatletter
\newcommand\hide@visible[1]{%
  \bgroup\fboxsep=.3ex\colorbox{Gray}{begin hide}%
  #1\colorbox{Gray}{end hide}\egroup%
}
\newcommand\hide@hidden[1]{%
  \bgroup\fboxsep=.3ex\colorbox{Gray}{hidden text}%
}
\newcommand\hide@invisible[1]{}
\newcommand\makevisible{\let\hide\hide@visible}
\newcommand\makehidden{\let\hide\hide@hidden}
\newcommand\makeinvisible{\let\hide\hide@invisible}
\makeatother
\makehidden

\usepackage{etoolbox}
\let\bbordermatrix\bordermatrix
\patchcmd{\bbordermatrix}{8.75}{4.75}{}{}
\patchcmd{\bbordermatrix}{\left(}{\left[}{}{}
\patchcmd{\bbordermatrix}{\right)}{\right]}{}{}

\begin{document}


\title{Glass transitions and scaling laws within an\\
alternative mode-coupling theory}






\author{
        Wolfgang G\"otze$^1$
          and Rolf Schilling$^2$\\[\baselineskip]%
                 \textit{$^1$ Physik Department, Technische Universit\"at M\"unchen,}\\
                 \textit{James-Franck-Str. 1, D- 85747 Garching, Germany}\\
                 \textit{$^2$ Institut f\"ur Physik, Johannes Gutenberg-Universit\"at,}\\
                \textit{Staudinger Weg 9, D-55099 Mainz, Germany}}
\date{\today}

\begin{abstract}
Idealized glass transitions are discussed within a novel mode-coupling theory (TMCT) proposed by Tokuyama(Physica A{\bf 395},31(2014)). This is done in order to identify common grounds with and differences to the conventional mode-coupling theory (MCT). It is proven that both theories imply the same scaling laws for the transition dynamics, which are characterized by two power-law decay functions and two diverging power-law time scales. However, the values for the corresponding anomalous exponents calculated within both theories differ from each other. It is proven that the TMCT, contrary to the MCT, does not describe transitions with continuously vanishing arrested parts of the correlation functions. It is also demonstrated for a schematic model that the TMCT neither leads to the MCT scenarios for transition-line crossings nor for the appearance of higher-order glass-transition singularities.
\end{abstract}





\maketitle


\section{Introduction}
With decreasing temperature $T$ or increasing density $\rho$ the dynamics of a liquid slows down drastically, and finally it arrests in a glass state. This holds provided crystallization can be avoided. An outstanding feature of glassy dynamics is stretching: the relaxation spectra extend over many orders of magnitude for the frequency variations. One cannot describe meaningfully the relaxation functions as superpositions of a number of exponential decay functions. Mode-coupling theory (MCT) is an attempt to describe the indicated features within the microscopic theory of simple liquids \cite{1,2}. The basic quantity of this approach is the intermediate scattering function $S(q,t)=\langle \rho(\vec{q}, t)^* \rho (\vec{q})\rangle/N$. For an N-particle system it describes the variations of the correlations with increasing time $t$ of density fluctuations $\rho(\vec{q}, t)$ of wave vector $\vec{q}$, $|\vec{q}|=q$. The equation of motion reads
\begin{equation}\label{eq1a}
\ddot{S}(q,t) + \nu_q \dot{S}(q,t)+\Omega^2_q [S(q,t) + \int^t_0 m_q (t-t') \dot{S}(q,t') dt'] =0 \ . 
\end{equation}
It is to be solved for the initial conditions $S(q,t=0)= S_q$ and $\dot{S}(q,t=0)=0$. This formula generalizes a Navier-Stokes-equation result by replacing the sound frequency by $\Omega_q$ and by complementing the friction term $\nu_q$ by a retardation term. The latter is quantified by a fluctuating-force kernel $m_q(t)$ \cite{3,4}. Here $S_q$ denotes the static structure factor, which depends smoothly on $T$ and $\rho$. Often it is more convenient to discuss dynamics in the domain of complex frequencies $z$, $Im(z) > 0$. Eq.~(\ref{eq1a}) is equivalent to the double-fraction formula

\begin{equation}\label{eq1b}
\hat{S}(q,z)/S_q=-1/[z-\Omega^2_q /[z + i \nu_q + \Omega^2_q \hat{m}_q(z)]]  \ .  
\end{equation}

Here and in the following Laplace transforms $\hat{F}(z)$ of functions $F(t)$ are defined by
\begin{equation} \label{eq2}
\hat{F}(z)=LT[F(t)](z)=i \int^\infty_0 F(t) \exp[izt] dt \quad.
\end{equation}
Eqs.~(\ref{eq1a}) and (\ref{eq1b}) are still exact, but with an unknown kernel $m_q(t)$ and $\hat{m_q}(z)$, respectively. To close the equation of motion, an ansatz for the kernel was motivated, which expresses $m_q(t)$ as a polynomial of $S(q,t)$. Details will be specified in Sec. IIA. Analogous MCT equations have been formulated for other functions, like the density-fluctuation correlator $S^{(s)}(q,t)$ for a tagged particle moving in the liquid  \cite{1,2}.

MCT describes ideal glass transitions occurring at some critical temperature $T_c$ or critical density $\rho_c$. For $T > T_c$ or $\rho < \rho_c$ the long-time limits of $S(q,t)$ vanish as expected for a liquid state. However, for $T \leq T_c$ or $\rho \geq \rho_c$, the correlations arrest for long times: $\lim_{t \rightarrow \infty} [S(q,t)/S_q] =\lim_{z \rightarrow 0} -[z \hat{S}(q,z)/S_q]=f_q$. The positive arrested parts $f_q$ are the Debye-Waller factors of an amorphous solid, i.e.~of a glass.

MCT also describes the evolution of a stretched dynamics as precursor of the ideal glass transition. Universal formulas have been derived in the limit $T$ tending to $T_c$ or $\rho$ approaching $\rho_c$. Two scaling laws describe stretching as interplay of two power-law-decay functions. The subtle dependence of the dynamics due to changes of the small parameters $(T-T_c)$ or  $(\rho_c-\rho$) is caused by two power-law-divergent time scales, which are specified by two anomalous exponents \cite{2}.

So far it has not been possible to control the shortcomings of the MCT ansatz for the kernel $m_q(t)$. This has challenged extensions and modifications of the MCT \cite{5,6,7,8,9,10}. Particularly stimulating is the work of Biroli and Bouchaud \cite{10}, which is built on equations for averages of products of four density fluctuations. They conclude that MCT is a mean-field approximation of their more general equations. But studies of the MCT for hard-sphere systems in high-dimensional spaces do not provide definite validity of this conclusion \cite{11,12}.

 Recently, an alternative to MCT has been studied by Tokuyama \cite{13} which will be called TMCT. While MCT is imbedded in the Zwanzig-Mori projection-operator formalism \cite{3,4}, the alternative TMCT is based on a different projection-operator formalism \cite{14}. Like in MCT also in TMCT an ansatz for a relaxation kernel $m_q(t)$ is required. It was suggested \cite{13} to use the same expression as was motivated in MCT. 

It was shown already that TMCT overestimates the glass formation less than MCT does \cite{15}. It was also proven that the non-Gaussian parameter for $t=0$ vanishes within TMCT while MCT yields a non-zero value \cite{13}. The aim of the present paper is to investigate within TMCT the glass transition scenarios and the long-time dynamics
and to compare with the corresponding results within MCT, thereby providing a motivation to test both theories by experiments and molecular-dynamics simulations. 

The present  paper is organized as follows. Sec. II presents the equations of motion for MCT and TMCT, reviews the basic concepts and manipulations in order to discuss the glass-transition singularities and it contains the  proof that TMCT - contrary to the MCT - does not exhibit continuous glass transitions. Sec. III and Sec. IV discuss within TMCT the validity of the first and second scaling law, respectively. The 
glass-transition diagram of a schematic model is calculated analytically within TMCT and its properties are compared with those calculated within MCT in Sec. V. The final section VI presents a summary and conclusions. In order to keep the presentation of our work self-contained the crucial concepts and steps will be explained within each section. Some technical details are put into an appendix.

\section{Glass-transition singularities}

In a first step, the equations of motion for MCT and TMCT will be presented in subsection A. Subsections B and C will describe necessary tools needed for the discussion of the glass-transition singularities and the dynamical features. Based on this, in subsection D it will be proven that TMCT can not describe continuous glass transitions.

\subsection{Equations of motion}

The dynamics shall be described by the normalized correlation functions $\phi_q(t)=S(q,t)/S_q$. The wave numbers shall be discretized to a set of $M$ values for the moduli $q_1, \cdots, q_M$. It is the goal to evaluate the array
$\bm{\mathcal{ \phi}} (t)=(\phi_1(t), \cdots, \phi _k (t), \cdots, \phi_M (t))$,  where $k$ corresponds to $q_k$. This shorthand notation will be used throughout the following. As mentioned above, arrays of kernels ${\bf m}(t)$ have to be defined. This is done by introducing arrays ${\bm{\mathcal{F}}}[{P, {\bf x}}]$ of mode-coupling polynomials:
\begin{equation} \label{eq3}
\mathcal{F}_q [P, {\bf x} ] =\sum^{n_0}_{n=1} \sum_{k_1 \cdots  k_n}
\, V^{(n)}_{q, k_1 \cdots k_n} (P) x_{k_1} \cdots x_{k_n} \quad .
\end{equation}
$P=(P_1, \cdots , P_d)$ denotes a point in the d-dimensional manifold of control parameters. These points specify the equilibrium state of the system. The coefficients $V^{(n)}_{q, k_1 \cdots k_n} (P)$ play the role of the coupling constants. They are non-negative smooth functions of the control parameters. The ansatz for the mode-coupling kernel reads

\begin{equation} \label{eq4}
{\bf m} (t) ={\bm{\mathcal{F}}} [P, \bm{\mathcal{ \phi}} (t) ] \quad .
\end{equation}
For studies of simple liquids quadratic polynomials have been motivated and the $V^{(2)}_{q, kp} (P)$ are expressed by $S_q$ \cite{1}. For a study of a spin-glass-transition model only coefficients for $n=1$ and $n=3$ have been used \cite{16}. Coefficients for $n=2$ had to be ignored because of symmetry reasons. In a study of the percolation problem for a hard sphere moving in a random arrangement of point scatterers only linear mode-coupling polynomials occur, i.e.~$V^{(n)}_{q, k_1 \cdots k_n} (P)=0$ for $n > 1$ \cite{17}. 
Equation~(\ref{eq3})
represents a generalized mode-coupling polynomial with a lower and an upper cut-off for $q$ at $q_1$ and $q_M$, respectively,  and with discretized wave numbers. Since, 
e.g. for a liquid, the  coefficients $V^{(n)}_{q, k_1 \cdots k_n} (P)$  involve the direct correlation functions, they decay for the wavenumbers going to infinity. This decay is such that for $M$ and  $q_M$ large enough the solutions of the MCT equations are practically independent of $M$ and $q_M$. The properties of a discontinuous glass transition are also robust under the continuum limit of the wave vectors, including the limit  $q_1 \to 0$. However, for  continuous transitions, e.g. for the tagged particle dynamics, the continuum limit is more subtle, since it also involves the limit $q_1 \to 0$. In Ref.\cite{17a} it was proven that in three spatial dimensions the transition survives in that limit.  But, the transition point and properties close to it, e.g. the dependence of the arrested parts of the corresponding correlators on the distance to the transition point, depend weakly on $q_1$. In two dimensions the transition point, i.e. the critical density,  shifts to zero  for $q_1 \to 0$.

The MCT, as well as the TMCT, are based on two equations of motion. The first one is shared by both theories. It expresses a current correlator ${\bf{\hat{K}}}(z)$ in terms of the Laplace transform of ${\bf m}(t)$ as Zwanzig-Mori fraction:

\begin{equation} \label{eq5}
\hat{K}_q (z)=-\Omega^2_q / [z + i \nu_q +\Omega^2_q \hat{m}_q (z)] \quad.
\end{equation}
The frequencies $\Omega_q$ and $\nu_q$ specify the transient dynamics.

The second equation of motion formulates a connection between the density-fluctuation correlators and the current correlators. The MCT uses again a Zwanzig-Mori fraction:
\begin{subequations}\label{equations}
\begin{align}
  \label{eq6a}
  MCT: \ \hat{\phi}_q (z)=-1/[z + \hat{K}_q (z)] \ . \quad \quad \quad \quad \quad 
 \intertext{Combining Eqs.~(\ref{eq5}) and (\ref{eq6a}) one gets Eq.~(\ref{eq1b}). For the TMCT one appeals to an alternative projection-operator formalism \cite{14} in order to get a connection between ${\bf K} (t)$ and $\bm{\mathcal{ \phi}}(t)$ in form of a first order differential equation \cite{13}: $\dot{\phi}_q(t) +[ \int_0^t K_q (t') dt'] \phi_q(t)=0$. The latter has to be solved for the initial condition $\phi_q(t=0)=1$. One gets $-\ln(\phi_q(t))=E_q(t):= \int_0^t (t-t')K_q (t') dt'$, i.e.~}
 \label{eq6b}
 TMCT: \   \phi_q(t)=\exp(-E_q(t));  \ \hat{E}_q (z) =- \hat{K}_q(z)/z^2 \ . \quad 
\end{align}
\end{subequations}

Two comments on the basic equations of both theories might be added. First, in order to use a unified description for MCT and TMCT  notations have been changed relative to the preceding literature on TMCT.  This concerns the damping constant and the memory kernel.  The microscopic frequency term(first term on the r.h.s. of Eq.(2) in Ref.\cite{15}) has been  denoted by $\Omega_{q}^2$, which is equal to $K_q(t=0)$. The major point, however,  is that the function $K_{c}(q,t)$ from Eq.~(2) of Ref.\cite{15}  has been denoted by $E_q(t)$. 
Taking this into account,  the Laplace transform of that equation is 
identical to our Eq.~(\ref{eq5}), and Eq.(1) in Ref.\cite{15} coincides with  Eq.~(\ref{eq6b})  of the present contribution.  Second, since both theories are based on different projection formalisms, the current correlator $K_q(t)$ in MCT and TMCT are different functions.

\subsection{Fixed-point equations}

As mentioned in the introduction, glass states $P$ are characterized by the arrest of density-fluctuation correlations: ${\bf f}=\lim_{t \rightarrow \infty} \bm{\mathcal{ \phi}}(t) = \lim_{z \rightarrow 0} [-z \bm{\mathcal{\hat{ \phi}}} (z)]$. Equation~(\ref{eq4}) implies arrest of the mode-coupling kernels: $\bm{\mathcal {F}} [P, {\bf f}]=\lim_{t \rightarrow \infty} {\bf m}(t)= \lim_{z \rightarrow 0} [-z {\bf\hat{m}} (z)]$. Formula (\ref{eq5}) leads to an equation for the low-frequency asymptote of the current correlators:

\begin{equation} \label{eq7}
\lim_{z \rightarrow 0} [\hat{K}_q(z)/z]=1/\mathcal{F}_q [P, {\bf f}] , \quad q=1, \cdots
, M \ .
\end{equation}
Substitution  of this formula into Eq.~(\ref{eq6a}) yields $f_q=1/[1+ 1 / \mathcal{F}_q [P, {\bf f}]]$. Using Eqs.~(\ref{eq6b}) and (\ref{eq7}) one arrives at $f_q=\exp[\lim_{z \rightarrow 0} [z \hat{E}_q (z)]]=\exp [-\lim_{z \rightarrow 0} [\hat{K}_q (z)/z]] = \exp[-1/\mathcal{F}_q [P, {\bf f}]]$. Defining mappings $\mathcal{T}_q[P, {\bf x}]$ of the manifold of arrays ${\bf x}$ into itself by

\begin{subequations}\label{equations}
\begin{align}
  \label{eq8a}
  MCT:  \ \mathcal{T}_q[P, {\bf x}]=1/[1+1/\mathcal{F}_q [P, {\bf x} ]] \ , \\
    \label{eq8b}
      TMCT: \  \mathcal{T}_q [P, {\bf x}] =\exp[-1/\mathcal{F}_q[P, {\bf x}]]  \ , 
\end{align}
\end{subequations}
the arrested parts ${\bf f}$ are fixed points of $\mathcal{T}_q[P, {\bf x}]$:

\begin{equation} \label{eq9}
{\bf f} = \bm{\mathcal{T}} [P, {\bf f}] \quad .
\end{equation}

For the manifold of arrays a semi-ordering can be defined by: ${\bf x} \geq {\bf y}$ if $x_q \geq y_q$, $ q=1, \cdots, M$. Introducing the special arrays ${\bf 0}=(0, \cdots, 0)$ and ${\bf 1}=(1, \cdots, 1)$ one can write ${\bf 1} > {\bf f} \geq{\bf 0}$. There hold the monotony properties: if ${\bf x} \geq {\bf y} \geq{\bf 0}$ one gets $\bm{\mathcal{F}} [P, {\bf x}] \geq \bm{\mathcal{F}}[P, {\bf y}]$ and $\bm{\mathcal{T}} [P, {\bf x}] \geq
\bm{\mathcal{T}} [P, {\bf y}]$. A sequence  of arrays ${\bf g}^{(n)}$, $n=0,1, \cdots$ can be defined recursively by: ${\bf g}^{(0)}={\bf 1}$, ${\bf g}^{(n+1)}= \bm{\mathcal{T}} [P, {\bf g}^{(n)}]$. Since ${\bf g}^{(n)} \geq {\bf g}^{(n+1)} \geq{\bf 0}$, the sequence converges towards some fixed point ${\bf f}^* \geq{\bf 0}$.

\subsection{The stability matrix}

The analysis in this subsection follows that perfomed for MCT. For more details the reader may consult Secs. 4.3.3 and 4.3.4 of Ref.\cite{2}.
The fixed-point equation Eq.~(\ref{eq9}) formulates $M$ implicit relations $\mathcal{I}_q [P, {\bf f}]=f_q-\mathcal{T}_q [P, {\bf f}]=0$, \ $q=1, \cdots , M$ for the $M$ components of a fixed point ${\bf{f}}(P)$. Let us assume that $P^*$ is a glass state specified by the fixed point ${\bf{f}}(P^{*})={\bf f}^*$. There appears the question whether this special solution can be extended to a neighborhood of states $P$ ``near'' $P^*$. The implicit-function theorem answers this question affirmatively provided the $M$-by-$M$ matrix with elements $\mathcal{I}_{q,k}(P^*, {\bf f}^*)= \partial \mathcal{I}_q/ \partial x_k[P^*, {\bf{f}^*}]$ can be inverted. Writing $\mathcal{I}_{q,k} [P, {\bf x}]= \delta_{q,k} - \tilde{A}_{q,k} (P, {\bf x})$ with $\tilde{A}_{q,k} [P, {\bf x}]=\partial \mathcal{T}_q [P, {\bf x}]/ \partial x_k$ one concludes the following. If the matrix $(\tilde{A}_{q,k} [P, {\bf x}])$ does not have an eigenvalue unity, a smooth continuation  ${\bf f}(P)$
of the fixed point array ${\bf f}^*$ is defined uniquely for a whole neighborhood of states $P$ near $P^{*}$. The eigenvalue condition for $(\tilde{A}_{q,k} [P, {\bf x}])$ can be replaced by one for matrices $(A_{q,k} [P, {\bf x}])=(s_q({\bf x})^{-1} \tilde{A}_{q,k} [P, {\bf x}]s_k(\bf x))$, $s_q({\bf x}) > 0$, since both matrices are connected by a similarity transformation. The choice of ${\bf{s}}(\bf x)$ is a matter of convention. Here we will choose $s_q(\bf x)$ such that $s_q({\bf{f}}(P)) \mathcal{F}_q[P,{\bf{f}}(P)]=1$. This choice simplifies the calculations in Sec. III. The new matrix $(A_{q,k} [P, {\bf x}])$ is called the  stability matrix of the theory.

From Eqs.~(\ref{eq8a}) and (\ref{eq9}) one gets $\tilde{A}_{q,k} [P, {\bf x}]=(1-x_q)^{2}\partial \mathcal{F}_q [P, {\bf x}]/ \partial x_k$ and  Eqs.~(\ref{eq8b}) and (\ref{eq9}) yield $\tilde{A}_{q,k} [P, {\bf x}] = x_q (\ln (1/x_q))^{2}
\partial \mathcal{F}_q [P, {\bf x}]/ \partial x_k$. Use of Eqs.~(\ref{eq8a}), (\ref{eq8b}), (\ref{eq9}) and the above convention for $s_q(\bf x)$ leads to

\begin{subequations}\label{equations}
\begin{align}
  \label{eq10a}
  MCT:  \  s_q({\bf{x}}) = (1-x_q)/x_q \ , \ \ \ \ \\
    \label{eq10b}
      TMCT: \  s_q({\bf{x}}) = \ln (1/x_q)  \ .  \ \ \ \quad 
\end{align}
\end{subequations}
Then the stability matrices  are given by

\begin{subequations}\label{equations}
\begin{align}
 \label{eq10c}
   MCT:  \  A_{q,k} [P, {\bf x}] =x_q(1-x_q) 
\partial \mathcal{F}_q [P, {\bf x}]/ \partial  x_k (1-x_k)x^{-1}_k , \ \ \\
\label{eq10d}
 TMCT:  \  A_{q,k} [P, {\bf x}] =x_q \ln(1/x_q)  
\partial \mathcal{F}_q [P, {\bf x}]/ \partial  x_k \ln(1/x_k) \ . \ \
\end{align}
\end{subequations}
Generically, the $M^2$ elements of the stability matrix are
positive. Some consequences of this property shall be cited, which
will be needed in the following. The matrix has a non-degenerate
maximum eigenvalue to be denoted by $e(P)$. All other eigenvalues have a modulus smaller than $e(P)$. The implicit-function theorem permits the following conclusions. Let $P^*$ denote a state with a fixed point ${\bf f}^*$ and $e(P^*) < 1$. Then there exists a neighborhood  of states $P$ with fixed points ${\bf f} (P)$ and $e(P) < 1$. The $M$ components of ${\bf f}(P)$ and $e(P)$ are smooth functions of the control parameters and ${\bf f} (P^*)={\bf f}^*$. Such $P^*$ are called regular glass states.

Glass-transition singularities or critical states $P^c$  with arrested parts
${\bf f}^c={\bf f}(P^{c})$ are characterized by a critical stability matrix $(A^c_{q,k})=(A_{q,k} [P^c, {\bf f}^c])$ with maximum eigenvalue unity: $e(P^c)=1$. Let $\bf{a}^*$ and ${\bf a}$ denote the left and right eigenvector of $(A^c_{q,k})$, respectively:

\begin{subequations}\label{equations}
\begin{align}
  \label{eq11a}
  \sum_p a^*_p A^c_{p,k}=a^*_k \quad , \quad \sum_p A^c_{q,p} a_p=a_q  \ .
\intertext{Generically all components of the eigenvectors can be chosen positive:}
  \label{eq11b}
 a^*_k > 0 \, , \, a_q > 0 \quad, k,q=1, \cdots, M  \ \ .
\intertext{The eigenvectors shall be fixed uniquely by the conventions}
  \label{eq11c}
  \sum_{q} a^*_q a_q=1 \quad ,\quad \sum\limits_q a^*_q a_q a_q/f^c_q=1 \quad.
  \end{align}
\end{subequations}

In Secs. III and IV there will appear the problem to understand the solution $\bf X$ of a set of $M$ linear equations. These are specified by the critical stability matrix and by an array ${\bf I}$ of inhomogeneities:\\

\begin{subequations}\label{equations}
\begin{align}
  \label{eq12a}
  \sum_k [\delta_{q,k} - A^c_{q,k} ] X_k=I_q  \quad, \quad q=1, \cdots, M \ \  .
\intertext{The solubility condition for this problem reads}
    \label{eq12b}
 \sum_q a^*_q I_q=0  \ \  . \quad \quad \quad \quad  \quad \quad     
\intertext{The general solution of Eq.~(\ref{eq12a}) is the sum of two terms. The first one is an arbitrary multiple X of the components of ${\bf a}$. The second
one is a linear combination of the components of ${\bf I}$ with coefficients $R_{q,k}$ given by the elements of the stability matrix:} 
  \label{eq12c}
X_q=X a_q + \tilde{X}_q \quad ;  \  \tilde{X}_q=\sum_k R_{qk} I_k \quad .
  \end{align}
\end{subequations}

\subsection{Absence of continuous glass transitions within TMCT}

Let $P_\varepsilon, \ -\varepsilon_0 < \varepsilon < \varepsilon_0$, denote a path in the manifold of control parameters. It shall be a smooth one in the sense that all
mode-coupling coefficients $V^{(n)}_{q,k_1 \cdots k_n } (P_\varepsilon)$ are smooth functions of the path parameter $\varepsilon$. The convention shall be made: states for
$\varepsilon < 0$ are liquids, the ones for $\varepsilon >0$ shall denote regular glass states, and $P_{\varepsilon=0}=P^c$ is a glass-transition singularity. It will be the goal
of this subsection to prove the theorem:
\vspace{0.2cm}

\textit{The TMCT equations (\ref{eq8b}) and (\ref{eq9}) with the mode-coupling polynomial (\ref{eq3}) cannot yield arrested parts ${\bf f}(P_{\epsilon})$  which change continuously with changes of $\epsilon$ at $P_{\epsilon=0}=P^c$}.

\vspace{0.2cm}

The arrays ${\bf x}$ which enter the mode-coupling polynomial obey the restriction ${\bf 1} \geq {\bf x} \geq {\bf 0}$. There is only a finite number of terms $V^{(n)}_{q, k_1 \cdots k_n} \, x_{k_1} \cdots x_{k_n}$ which enter Eq.(\ref{eq3}). Hence there is a finite bound $L'$  so that

$$\sum\limits_{k_1 \cdots k_n} V^{(n)}_{q, k_1 \cdots k_n} x_{k_1} \cdots x_{k_n} < L' \sum\limits_{k_1 \cdots k_n} x_{k_1} \cdots x_{k_n} \quad.$$
The latter sum also contains a finite number of terms. Therefore a bound $L''$ exists such that $\sum\limits_{k_1 \cdots k_n} x_{k_1} \cdots x_{k_n} \leq L'' \sum\limits_k x_k$. Consequently, there is a finite positive constant $L$ so that the mode-coupling polynomial is bounded by a linear one:

 \begin{subequations}\label{equations}
\begin{align}
  \label{eq13a}
  \mathcal{F}_q[P_\varepsilon, {\bf x}] \leq L \ \xi ({\bf x}) \quad , \quad \xi({\bf x}) =\sum\limits_k x_k \ , \ \ \ \ \ \ \ \quad
\intertext{for all $q$. Substituting this inequality into the TMCT map, Eq.~(\ref{eq8b}), one finds $\sum\limits_{q} \mathcal{T}_q[P_\varepsilon, {\bf x}] \leq M \exp[-1/(L \xi ({\bf x}))]$.
Elementary considerations show the following. There is a positive number $\xi_{\rm max}$ such that for all $\xi < \xi_{\rm max}$ the inequality $M \exp [-1/(L \xi)] < \xi \exp (-1/L)$ is valid. This leads to}
    \label{eq13b}
  \sum\limits_q \mathcal{T}_q [P_\varepsilon, {\bf x}]\leq \xi ({\bf x}) \exp(-1/L) \, ,\quad 0 < \xi({\bf x}) < \xi_{\rm max}\, \ .
    \end{align}
\end{subequations}
The inequality (\ref{eq13b}) is the basis for the proof of the theorem.
The proof is done indirectly. Let us assume that there is a state
$P_\varepsilon$ with an associated fixed point ${\bf f}^\varepsilon=
{\bf f} (P_\varepsilon)$ which  tends continuously to zero for
$\varepsilon \rightarrow 0$. In particular
$\xi_\varepsilon=\sum\limits_q f^{\epsilon}_q$ tends continuously to zero. Consequently, one can find a state $P_\varepsilon$ close to $P^c$ such that $\xi_\varepsilon < \xi_{\rm max}.$ However, this result contradicts the inequality (\ref{eq13b}) since $\exp(-1/L)< 1$ and $\sum\limits_q \mathcal{T}_q [P_\varepsilon, {\bf f}^\varepsilon] = \xi_\varepsilon$, which follows from the fixed-point equation (\ref{eq9}).

\section{The first scaling law}

In this section the TMCT equations of motion shall be reformulated for a treatment of states $P_\varepsilon$ with $\varepsilon$ tending to zero. This is done as originally performed for MCT
(see Secs. 6.1.1 and 6.1.3 of Ref. \cite{2}).
The remaining discussions shall be restricted to the TMCT.

In the limit of $\varepsilon \rightarrow 0$ the states $P_\varepsilon$ tend towards the glass-transition singularity $P^c$. A first ansatz shall be made by treating $(\phi_q(t)-f^c_q)$ as a small quantity of order $\sqrt{|\varepsilon|}$. The equations shall be reformulated up to terms of order $|\varepsilon|$. Terms of higher order shall be indicated by the symbol $\mathcal{O}_\varepsilon$, i.e.~$\lim_{\varepsilon \rightarrow 0} \mathcal{O}_\varepsilon/\varepsilon=0$. Let us write

\begin{subequations}\label{equations}
\begin{align}
  \label{eq14a}
 \phi_q(t)-f^c_q=s^c_q \ g_q(t) \ , \\
    \label{eq14b}
   s^c_q= \ln(1/f^c_q)    \ ,
\end{align}
\end{subequations}
where the amplitude $s^c_q$ follows from  Eq.~(\ref{eq10b}) for $\bf x= {\bf f}^c$
The ansatz implies

\begin{subequations}\label{equations}
\begin{align}
  \label{eq15a}
  g_{k_1} (t) \cdots g_{k_n} (t) = \mathcal{O}_\varepsilon \quad, \quad n \geq 3 \quad.
\intertext{For liquid states, the ansatz is non-trivial since restrictions of $g_q(t)$ to order $\sqrt{|\varepsilon|}$ imply restrictions of the admissible time interval
for small as well as for large $t$. A second ansatz concerns the regular term in Eq.~(\ref{eq5}). It will be assumed that it can be neglected relative to the mode-coupling kernel \ :}
    \label{eq15b}
  [z + i \nu_q]/[\hat{m}_q (z)]=\mathcal{O}_\varepsilon \quad.
      \end{align}
\end{subequations}
The two preceding formulas shall be justified a posteriori at the end of this section in connection with the definition of a scaling limit.

The reformulations start by substitution of Eq.~(\ref{eq14a}) into Eqs.~(\ref{eq3}) and ~(\ref{eq4}) in order to obtain the relaxation kernel as a polynomial in ${\bf g} (t):$

\begin{subequations}\label{equations}
\begin{align}
  \label{eq16a}
s^c_q m_q(t)=1+ \sum_k W^{(1)}_{q,k} g_k(t) + \ \ \ \  \ \ \ \ \ \ \ \ \ \nonumber\\
+ \sum_{kp} W^{(2)}_{q,kp} g_k(t) g_p(t) + D_q [P_\varepsilon] + \mathcal{O}_\varepsilon \ . \ \ \ \ \ \quad
\intertext{The convention made in Eq.~(\ref{eq14b}) implies that the leading order of the expansion of $s^c_q m_q(t)$ is unity. The first order coefficients are related to the critical stability matrix:}
 \label{eq16b}
W^{(1)}_{q,k} = s ^c_q \partial \mathcal{F}_q/\partial f_k [P^c, \, {\bf f}^c] s^c_k  =[1/f^c_q] A^c_{q,k} \ . \   
\intertext{The second order coefficients read:}
\label{eq16c}
W^{(2)}_{q,kp} =s ^c_q \partial^2 \mathcal{F}_q/\partial f_k \partial f_p  [P^c, \, {\bf f}^c] s^c_k s^c_p/2 \ .  \   \quad
\end{align}
\end{subequations}
The coefficients are to be evaluated at the critical point. Smooth variations with $\varepsilon$ can be absorbed in $\mathcal{O}_\varepsilon$. The dependence on control parameters enters via the smooth functions of $\epsilon$.

\begin{equation} \label{eq17}
D_q [P_\varepsilon]=s^c_q \{\mathcal{F}_q [P_\varepsilon, {\bf f}^c] - \mathcal{F}_q [P^c, {\bf f}^c] \} \quad.
\end{equation}
Further reformulations for $\hat{K_q}(z)$ and $E_q(t)$ are needed below and they are presented in an appendix.

Equation (\ref{A5}) together with the formulas (\ref{A6})~-~(\ref{A8}) for $I_q(t)$ provide  the reformulated equation of
motion for $g_q(t)$. How to solve such an inhomogeneous equation has been discussed in the last paragraph of Sec. IIC. The array ${\bf X}$ in
Eq.~(\ref{eq12a}) can be identified with the array ${\bf g} (t)$ in Eq.~(\ref{A5}). According to the ansatz it is of leading first order in the small
parameter $\sqrt{|\varepsilon|}$.
It contributes to Eq.~(\ref{eq12c}) the term $g^\varepsilon(t) {\bf a}$, where the factor $X$ in Eq.~(\ref{eq12c}) is denoted by $g^{\epsilon}{(t)}$. According to Eq.~(\ref{A6}), the array
${\bf I} (t)$ consists of three terms of second order in $\sqrt{|\varepsilon|}$. Thus, the contribution ${\tilde{X}_q}$ in Eq.~(\ref{eq12c}) does not contribute to the
leading-order results, to which the following discussions are restricted. Therefore Eq.~(\ref{eq14a}) takes the form

\begin{subequations}\label{equations}
\begin{align}
  \label{eq18a}
\bm{\mathcal{ \phi}} (t)= {\bf f}^c + {\bf h} g^\varepsilon (t) \quad,
\intertext{with the critical amplitudes introduced by}
\label{eq18b}
h_q=s^c_q a_q  \ , \ q=1, \cdots, M \ .
\end{align}
\end{subequations}

The preceding reasoning is correct if and only if the solubility condition Eq.~(\ref{eq12b}) is obeyed. The first contribution to $\sum_q a_q^* I_q(t)$ is a smooth
function which vanishes for $\varepsilon=0$. It is called the separation parameter $\sigma(\varepsilon)$:

\begin{equation} \label{eq19}
\sigma(\varepsilon) =\sum_q a^*_q f^c_q D_q [P_\varepsilon]= C \varepsilon \, , \, \ C > 0 \  .
\end{equation}
The second equality sign holds for generic paths, and if terms of order $(\sqrt{|\varepsilon|})^4$ are neglected. The second contribution yields $\lambda g^\varepsilon(t)^2$ with
$\lambda=\sum_{q, kp} a^*_q A^{c}_{q, k p} a_k a_p$. Substitution of Eq.~(\ref{A7}) and eliminating $W^{(1)}_{q,k}$ by using Eqs.~(\ref{eq16b}),~(\ref{eq11a}), and~(\ref{eq11c})
one gets the so-called exponent parameter

\begin{equation} \label{eq20}
\lambda=\sum_{q,kp} a^*_q f^c_q W^{(2)}_{q, kp} a_k a_p + (1/2) \sum_q s^c_q/f^c_q \, \ .
\end{equation}
It differs from the exponent parameter within MCT, particularly due to the presence of the second term.
The third contribution is a product $\lambda' LT ^{-1} [z \hat{g}^\varepsilon (z)^2] (t)$. Using Eqs.~(\ref{A8}), (\ref{eq16b}),~(\ref{eq11a})  and~(\ref{eq11c}) one gets $\lambda'= \sum_q a^*_q f^{c}_q W^{(1)}_{q,k} W ^{(1)}_{q,p} a_k a_p=1.$

As a result, the TMCT equations of motion are reduced to the scaling equation 

\begin{equation} \label{eq21}
\sigma(\varepsilon) + \lambda g^\varepsilon (t)^2 - \partial_t \int^t_0 g^\varepsilon (t-t') g^\varepsilon (t') dt'=0 \quad.
\end{equation}
This is the same non-linear integro-differential equation, which was studied in MCT. Only those implications shall be cited, which are needed  for the discussion below.

For $\sigma=0$, Eq.~(\ref{eq21}) is solved by the critical correlator $g^c(t)=(t_0/t)^a$. This holds provided the critical exponent $a$ is related to the exponent  parameter $\lambda$ according to:

\begin{equation} \label{eq22}
\Gamma(1-a)^2 / \Gamma (1-2a)=\lambda \quad, \quad 0 < a<1/2 \quad.
\end{equation}
The ansatz (\ref{eq15b}) has eliminated the effect of the transient dynamics on the solutions. Therefore, the time scale $t_0$ cannot be determined analytically. It has to be chosen such that
the critical correlator matches the transient. Let $g^\pm (\hat{t})$ denote the solution of Eq.~(\ref{eq21}) for $\sigma=1$ (glass) and for $\sigma=-1$ (liquid). The critical correlator shall be chosen as inital
condition\\
\begin{subequations}\label{equations}
\begin{align}
  \label{eq23a}
 \lim_{\hat{t} \rightarrow 0} g^\pm (\hat{t}) \hat{t}^a=1 \ . \ \quad \quad \quad \quad \quad \quad  
\intertext{Then a solution of the scaling equation Eq.~(\ref{eq21}) is given by:}
\label{eq23b}
g^\varepsilon(t) = c_\sigma g^\pm (\hat{t}) \quad , \quad \hat{g}^\varepsilon(z)=c_\sigma t_\sigma \hat{g}^\pm (\hat{z}) \, \ .
\intertext{Here rescaled times $\hat{t}$ and rescaled frequencies $\hat{z}$ are defined by}
\label{eq23c}
\hat{t}=t/t_\sigma \quad , \quad \hat{z}= z t_ \sigma \ .  \ \ \quad \quad \quad \quad \quad \quad
\intertext{This holds provided the correlation scale $c_\sigma$ and the time scale $t_\sigma$ are chosen as}
\label{eq23d}
c_\sigma=\sqrt{|\sigma|} \, , \ \  t_\sigma=t_0/|\sigma|^\delta \quad , \quad \delta =1/2 a \ \ .
\end{align}
\end{subequations}
The preceding equations formulate the \textit{first scaling law}. They describe the evolution of the dynamics in terms
 of control-parameter independent shape functions $g^\pm(\hat{t})$ or $\hat{g}^\pm (\hat{z})$. The sensitive
 changes, which occur if the states $P_\varepsilon$ approach the glass-transition singularity are described by the  variations of the two scales $c_\sigma$ and $t_\sigma$.

 It remains the task to justify the ansatz, Eq.~(\ref{eq15b}). Indeed, one gets $[z + i \nu_q]/\hat{m}_q(z)=[\hat{z}/t_\sigma
 + i \nu_q] /[ \hat{m}_q (\hat{z}) t_\sigma]= \mathcal{O}_{\epsilon}$, because $(1/t_\sigma)=(1/t_0) C|\varepsilon| \cdot |\sigma(\epsilon)|^\eta$, with $\eta=(\delta-1)>0.$
 Behind the derivation of the first scaling law, there is the scaling limit for $\sigma {>\atop<} 0:$

\begin{equation} \label{eq24}
\lim_{\varepsilon \rightarrow 0, 0< \hat{t}\,  {\rm fixed}} [\phi_q (\hat{t} t_\sigma)-f^c_q]/(h_q c_\sigma)= g^\pm(\hat{t}) \ , \ q=1, \cdots, M \ . \quad
\end{equation}
For $\varepsilon \rightarrow 0$, there appears via the correlation scale $c_\sigma$ the small parameter $\sqrt{|\varepsilon|}$.  This was anticipated by the ansatz, Eq.~(\ref{eq14a}). But, the time shifts towards infinity as $\hat{t}  t_\sigma$. The scaling law views the dynamics on scale $t_\sigma$.

The shape function for the glass exhibits arrest: $\lim_{\hat{t} \rightarrow \infty} g^+ (\hat{t})=1/ \sqrt{1-\lambda}$. The asymptotic law for the fixed point reads:

\begin{equation} \label{eq25}
f_q=f^c_q + h_q \sqrt{\sigma/(1-\lambda)} \quad, \quad \sigma \rightarrow 0+ \quad.
\end{equation}
The square-root variation is the signature of a fold bifurcation. The divergency for $\lambda \rightarrow 1$ signalizes the approach to a higher-order
singularity. The $\phi_q(t)$-versus-$t$ curve describes the crossover from the critical decay to arrest.

The shape function for the liquid and $\hat{t} \rightarrow \infty$ exhibits power-law variation quantified by an exponent $b$:\\
\begin{subequations}\label{equations}
\begin{align}
  \label{eq26a}
 g^- (\hat{t}) \sim \ - B \hat{t} ^b    \ , \  {\hat{t} \rightarrow \infty} \  \ .  \quad \quad\quad 
\intertext{This is von Schweidler's law. The von Schweidler exponent $b$ is related to the exponent parameter $\lambda$ \ :}
\label{eq26b}
\lambda=\Gamma (1+b)^2 /\Gamma (1+2b) \quad, \quad 0< b \leq 1 \quad .
\end{align}
\end{subequations}
The constant $B$ cannot be calculated analytically. Substitution of these results into Eq.~(\ref{eq18a}) casts von Schweidler's law into the form:

\begin{subequations}\label{equations}
\begin{align}
  \label{eq27a}
\bm{\mathcal{ \phi}} (t) = {\bf f}^c - {\bf h} (t/t'_\sigma)^b   \ \ \quad \quad \quad \quad  \\
\label{eq27b}
t'_\sigma=t'_0 /|\sigma|^\gamma \quad, t'_0=t_0/B^{1/b} \, , \, \gamma=(1/2a)+ (1/2b) \, \, .
\end{align}
\end{subequations}
It describes power-law decay below ${\bf f}^c$. There appears a new time scale $t'_\sigma$ with $t'_\sigma/t_\sigma \propto 1/|\sigma|^{1/2 b}$

\section{The second scaling law}

In this section all considerations are restricted to liquid states close to the transition point: $\sigma < 0$, $P_\varepsilon \rightarrow P^c$. Validity of the second
scaling law of MCT will be proven for TMCT. Again, only the crucial steps will be given. For more details see Secs. 6.2.1 and 6.2.2 of Ref. \cite{2}. 

\subsection{The equations of motion}
A second scaling limit shall be analyzed, again justified a posteriori by its success. It is based on the introduction of a $\sigma$-sensitive time scale $\tau$, which
diverges more strongly than $t_\sigma: \lim_{\varepsilon \rightarrow 0} t_\sigma/ \tau =0$. Rescaled times and rescaled frequencies are introduced: $\tilde{t}=t/\tau$, $\tilde{z}=z\tau$. The
ansatz for the searched for array reads $\bm{\mathcal{ \phi}} (t)={\bf F} (\tilde{t}, \sigma)=\bm{\mathcal{\tilde{\phi}}}(\tilde{t})+\sigma \bm{\mathcal{\varphi}}(\tilde{t})+ \mathcal{O}_{\epsilon}$.
The scaling limit concerns the approach of $\sigma$ to zero for fixed $\tilde{t}$ and $\tilde{z}$. The leading-order result, to which the discussions shall be restricted, reads

\begin{equation} \label{eq28}
\bm{\mathcal{ \phi}} (t)= \bm{\mathcal{\tilde{\phi}}} (\tilde{t}) \quad ,
\quad \bm{\mathcal{\hat{\phi}}} (z)=\tau \bm{\mathcal{\hat{\tilde{\phi}}}} (\tilde{z}) \quad.
\end{equation}
The correlators are given by shape functions $\bm{\mathcal{\tilde{\phi}}}$ which are independent of the control parameters. The change of the dynamics due to changes of the state $P_\varepsilon$ is
caused solely by changes of the scale $\tau$. Substitution of $\bm{\mathcal{\phi}}(t)$ from Eq.~(\ref{eq28}) into Eq.~(\ref{eq4}) for the mode-coupling kernels implies the scaling law:

\begin{equation} \label{eq29}
{\bf m} (t)= {\bf{\tilde{m}}} (\tilde{t})=\bm{\mathcal{F}} [P^c, \bm{\mathcal{\tilde{\phi}}} (\tilde{t})] \quad.
\end{equation}
Changes of the coefficients $V^{(n)}_{q, k_1 \cdots k_n} (P_\varepsilon)$ in Eq.~(\ref{eq3}) due to changes of $P_\varepsilon$ would contribute only to
the corrections $\bm{\mathcal{\varphi}} (\tilde{t})$.

The first step towards closed relations of the rescaled quantities is based on Eq.~(\ref{eq5}). It reads $\hat{K}_q(z)=-(1/\tau)/\{( \tilde{z}/\tau+ i\nu_q)/(\tau \Omega^2_q) + {\hat{\tilde{m}}}_q (\tilde{z}) \}$.
Here we used  ${\bf \hat{m}}(z)=\tau {\bf \hat{\tilde{m}}}(\tilde{z})$, which follows from Eq.~(\ref{eq29}). For the scaling limit $\tau \rightarrow \infty$, but $\tilde{t}$ fixed, one gets\\
\begin{subequations}\label{equations}
\begin{align}
  \label{eq30a}
\tau \hat{K}_q(z) = {\hat{\tilde{K}}}_q(\tilde{z}) =-1/ {\hat{\tilde{m}}}_q (\tilde{z}) \quad.
\intertext{The second step deals with Eq.(\ref{eq6b}). One gets from Eq.~(\ref{eq6b}) with Eq.~(\ref{eq30a})}
\label{eq30b}
\hat{E}_q(z)=\tau/ [{\hat{\tilde{m}}}_q (\tilde{z}) \tilde{z}^2]=\tau {\hat{\tilde{E}}}_q (\tilde{z}) \  , \
\intertext{which implies}
\label{eq30c}
E_q(t)=\tilde{E}_q(\tilde{t})=L T^{-1} [1/[{\hat{\tilde{m}}}_q (\tilde{z}) \tilde{z}^2 ](\tilde{t}) \quad.
\end{align}
\end{subequations}
Hence, $\phi_q(t)=\exp[-E_q(t)]=\exp[-\tilde{E}_q(\tilde{t})] = \tilde{\phi}_q (\tilde{t})$. This proves
consistency of this procedure and it establishes the \textit{second
  scaling law}. The scale $\tau$ cannot be fixed after the
parameters $\Omega_q$ and $\nu_q$ for the transient effects have been eliminated.

The initial values of the shape functions, say, $f^{(i)}_q={\bf{\tilde{\phi}}}_q (\tilde{t} \rightarrow 0)$, are implications of the preceding equations, which do not depend on the scale $\tau$. One gets using Eqs.~(\ref{eq6b}),~(\ref{eq30b}) and~(\ref{eq29}):

$f^{(i)}_q=\exp [-\tilde{E}_q (\tilde{t} \rightarrow 0)] =\exp[\lim_{\tilde{z} \rightarrow \infty} [\tilde{z} {\hat{\tilde{E}}}_q
(\tilde{z})]]=\exp[\lim_{\tilde{z} \rightarrow \infty} [1/({\hat{\tilde{m}}}_q (\tilde{z}) \tilde{z})]]=\exp[-1/ \tilde{m}_q
(\tilde{t} \rightarrow 0)]= \exp[-1/\mathcal{F}_q[P^c, {\bf f}^{(i)}]].$
Equations (\ref{eq8b}) and~(\ref{eq9}) imply that ${\bf f}^{(i)}$ is a fixed
point for $P=P^c$, Generically, one gets

\begin{equation} \label{eq31}
\bm{\mathcal{\tilde{\phi}}} (\tilde{t} \rightarrow 0)= {\bf{f}}^c \quad .
\end{equation}
The second scaling law does not deal with the decay of correlations below unity, but it deals with the decay below $f^c_q$.

\subsection{von Schweidler's law}

 Equation~(\ref{eq31}) suggests to solve the equations of motion for the limit $\tilde{t}$ tending to zero. Writing

 \begin{equation} \label{eq32}
 \tilde{\phi}_q (\tilde{t})=f^c_q + s^c_q \tilde{g}_q (\tilde{t})\,
 \end{equation}
 the solution can be obtained using $\tilde{g}_q(\tilde{t})$ as small quantities. Substituting these expressions
 into Eq.~(\ref{eq29}) yields the expansion up to terms in second order:

 \begin{equation} \label{eq33}
 s^c_q \tilde{m}_q (\tilde{t})=1 + \sum_k W^{(1)}_{q,k} \tilde{g}_k (\tilde{t}) +\sum_{kp} W^{(2)}_{q,kp}\tilde{g}_k (\tilde{t})
 \tilde{g}_p (\tilde{t}) \quad.
 \end{equation}
The coefficients $W^{(1)}_{q,k}$ and $W^{(2)}_{q, kp}$ are given in Eqs.~(\ref{eq16b}) and~(\ref{eq16c}).
This expansion in terms of $\tilde{g}_q (\tilde{t})$ is identical to that given in Eq.~(\ref{eq16a}). Here,
the procedure is simplified since the term $D_q[P_\varepsilon]$ does not occur. The further reformulations of the
equations of motion can be obtained following the ones explained in the appendix. One ends up with Eq.~(\ref{A5}) with
the inhomogeneity from Eq.~(\ref{A6}), but with $D_q[P_\varepsilon]=0$.

The further discussion follows that presented in Eqs.~(\ref{eq18a}), ~(\ref{eq18b}), ~(\ref{eq19}) and ~(\ref{eq20}) for the derivation of the scaling
equation~(\ref{eq21}). One finds in leading order ${\bf{\tilde{g}}} (\tilde{t})={\bf{a}} \tilde{g} (\tilde{ t})$, where
$\bf{a}$ is the right eigenvector of the critical stability matrix specified in Eqs.~(\ref{eq11a})-~(\ref{eq11c}). Function
$\tilde{g}(\tilde{t})$ is a solution of Eq.~(\ref{eq21}) specialized for $\sigma(\varepsilon)=0$. A solution consistent
with the requested short time behavior reads $\tilde{g} (\tilde{t})=-\tilde{t}^b$. The positive exponent $b$ is determined
by Eq.~(\ref{eq26b}), where the exponent parameter $\lambda$ is noted in Eq.~(\ref{eq20}). For small rescaled times
$\tilde{t}=(t/\tau)$, there holds von Schweidler's law: $\bm{\mathcal{\tilde{\phi}}} (\tilde{t}) ={\bf f}^c - {\bf h} \tilde{t}^b$.
Up to here the scale $\tau$ is not fixed. It is determined by the matching request. If, and only if, the long-time asymptote
on scale $t_\sigma$ is identical with the small-time asymptote on scale $\tau$ the second scaling law is a continuation of the first one.
Equation (\ref{eq27a}) requires that $\tau=t'_\sigma$.

\section{A Schematic model}

Some features of the results shall be demonstrated by considering a model, which is simplified with the mere intention to
reduce the numerical effort. First, the array of $M$ components are chosen for $M=1$, i.e.~, only a single function $\phi(t)$ is used. Second, the mode-coupling polynomial is chosen as

\begin{equation} \label{eq36}
m[P,x] = \upsilon_1 x + \upsilon_3 x^3 \quad.
\end{equation}
The two coupling coefficients $\upsilon_1$ and $\upsilon_3$ are used as control parameters. A state is presented as a point of the
first quadrant of the $\upsilon_1-\upsilon_3$ plane: $P=(\upsilon_1, \upsilon_3)$, $\upsilon_1 \geq 0$, $\upsilon_3 \geq 0$. This mode-coupling polynomial is a schematic version of the corresponding polynomial for a spin glass transition \cite{16}.

Within TMCT the fixed-point equation (\ref{eq9}) reads $\upsilon_1 + \upsilon_3 f^2=1/[f \ln (1/f)]$. The transition singularities are given by the request that the critical stability ``matrix'' from Eq.~(\ref{eq10d}) is unity: $\upsilon^c_1 +  3 \upsilon^c_3 f^{c2}=1/[f^c (\ln (1/f^c))^2]$. These two equations imply formulas for the critical point $P^c= (\upsilon^c_1, \upsilon^c_3)$ as function of the arrested parts $f^c$:

\begin{subequations}\label{equations}
\begin{align}
  \label{eq35a}
\upsilon^c_1= [3 G_1 (f^c)-G_2 (f^c)]/2 ,  \  \upsilon^c_3=[G_2 (f^c)-G_1 (f^c)]/[2 (f^{c})^{2}] \, ,     \\
\label{eq35b}
G_1(f^c)=1/[f^c s^c], \ \, G_2 (f^c) =1/[f^c (s^{c})^2] \, , \, s^c=\ln(1/f^c) \quad.
\intertext{The exponent parameter from Eq.~(\ref{eq20}) becomes}
\label{eq35c}
\lambda=s^c[2-(3/2)s^c]    \ . \ \ \quad \quad \quad \quad\quad \quad \quad\quad 
\end{align}
\end{subequations}

\begin{figure}
\includegraphics[angle=0,width=8cm]{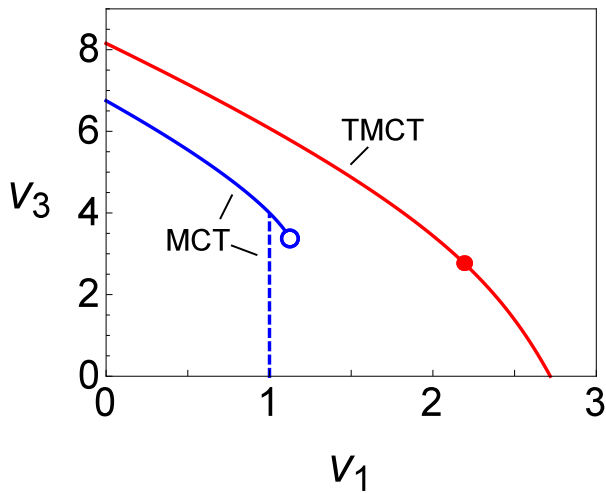}
\caption{(Colour online) Glass-transition diagram for a schematic model specified by a mode-coupling function $\mathcal{F}[P,x]=v_1x+v_3x^3$ within MCT \cite{18}
and TMCT. The dashed line marks the positions of continuous  and the solid lines represent positions of discontinuous transitions.  Both dots mark the state point $(\upsilon_1, \upsilon_3)$ with maximum exponent parameter $\lambda_{\rm max}=2/3$ for TMCT (full dot) and $\lambda_{\rm max}=1$ for MCT (open dot). The solid line for MCT terminates at the open dot.}
\label{fig1}
\end{figure}

The glass transition diagram calculated within TMCT
is presented in Figure~\ref{fig1}. The corresponding diagram calculated within MCT \cite{18} is added for reasons of comparison.
The figure demonstrates that the trend to glass formation is stronger in MCT than in TMCT, as already noted in Ref. \cite{15}. For $\upsilon^c_1=0$, the
TMCT value for $\upsilon^c_3$ exceeds that for the MCT by about 20\%. For $\upsilon^c_1=0$, both theories
yield $\lambda=1/2$, i.e. the von Schweidler exponent $b$ equals unity. For this case, there is no stretching for the decay which
is described by the second-scaling-law process.

Increasing $\upsilon^c_1$, the value $\upsilon^c_3$ decreases. Within MCT, this decrease is connected with an increase
of $\lambda$. The exponent parameter reaches unity at a higher-order glass-transition point(open dot in  Figure~\ref{fig1}). This increase is connected
with a decrease of the exponent $a$ and $b$ towards zero. The stretching  increases up to
the point, where the range of validity of both scaling laws shrinks to
zero. Within MCT, also a transition line(dashed line in  Figure~\ref{fig1}) occurs, which is connected
with a continuous variation of the arrested parts. This line terminates at a crossing point with the above mentioned transition line
of discontinuous changes of the arrested part. This crossing implies the existence of glass-to-glass transitions.

The TMCT does not exhibit any of the subtle phenomena described in the preceding paragraph. The $\upsilon^c_3$-versus-$\upsilon^1_c$ curve
decreases monotonously through the quadrant if $f^c$ decreases from $e^{-1/3}$ (at $\upsilon^c_1 =0$) to $e^{-1}$ (at $\upsilon^c_1=e)$.
With increasing $\upsilon^c_1$, the exponent parameter increases up to a maximum value $\lambda_{\rm max}=2/3$ (at  $\upsilon^c_1=9e^{2/3}/8)$ . This maximum value  is connected
with $f^c=e^{-2/3}$. For increasing $\upsilon^c_1$ further, $\lambda$ decreases. Consequently, there exists a minimal value of the von Schweidler
exponent $b_{\rm min} \approx 0.70$ corresponding to $\lambda_{\rm max}=2/3.$

\section{Summary and Conclusions}

We have elaborated on similarities and differences between the novel mode-coupling theory (TMCT), recently been proposed by
Tokuyama \cite{13}, and the conventional mode-coupling theory (MCT). Our focus has been on the glass-transition singularities and
the validity within TMCT of both scaling laws found within MCT.

A very important difference between both theories is the absence of a continuous glass transition within TMCT with mode-coupling polynomial (\ref{eq3}). This is related to the fact that all the  derivatives  $\partial \mathcal{T}_q [P, {\bf x}]/ \partial x_k$
vanish at ${\bf{x}=\bf{0}}$ within TMCT,  in contrast to MCT.

As an illustration the
phase diagram of a schematic model has been determined analytically. Figure 1 shows that there is a transition line within MCT at
which continuous glass transitions occur, in contrast to TMCT. This transition line terminates at  the MCT-transition line of discontinuous
transitions. Between the endpoint (open dot in Figure \ref{fig1}) of the latter transition line and the crossing point there occur glass-to-glass
transitions. These phenomena do not exist for TMCT. Another important difference is the absence of higher order glass-transition singularities for the schematic
model within TMCT, because the exponent parameter is always below unity. These results   challenge within TMCT investigations of microscopic models
like short-range attractive colloids. For such liquids MCT predicts higher order glass-transition singularities, glass-to-glass transitions and re-entrant
behavior \cite{19,20,21}.

The first scaling law found for MCT implies for the relaxation of the correlators two time power laws with diverging time scales. These involve exponents
which are determined by the exponent parameter $\lambda$. We have proven that
TMCT leads to the same scaling equation but with a different expression for $\lambda$. Consequently, the long-time relaxation obtained within TMCT
also exhibits two power laws with diverging time scales, however, with different numerical values for the corresponding exponents. Explicit formulas for the evaluation of the exponent parameter and for the amplitudes of the singular parts of the correlators have been obtained.
The second scaling law within MCT proves the existence of the superposition principle, e.g.~for the time and temperature dependence of a correlator. This
scaling law also holds within TMCT.

The kind of scaling laws originally found within MCT were not known before in the field of "Nonlinear dynamics", where bifurcations of dynamical states are investigated. The validity of both scaling laws within MCT and TMCT proves some  robustness of the dynamical features of the bifurcation scenario which describes the ideal liquid-to-glass transition.

\begin{acknowledgments}
We are very grateful to Michio Tokuyama for various discussions on properties of his alternative mode-coupling theory
\end{acknowledgments}

\begin{appendix}
\section{Reformulation of the TMCT equations of motion}
\renewcommand{\theequation}{\thesection\arabic{equation}}
\setcounter{equation}{0}
Laplace transformation of Eq.~(\ref{eq16a}) leads to

\begin{eqnarray} \label{eqA1}
&&-s^c_q z \hat{m}_q (z)=1-z \sum_k W^{(1)}_{q,k} \hat{g}_k (z) + D_q [P_{\varepsilon}]\nonumber\\
&& -z\sum_{kp} W^{(2)}_{q,kp} L T [g_k (t) g_p (t)] (z) + \mathcal{O}_\varepsilon    \ .
\end{eqnarray}

The ansatz in Eq.~(\ref{eq15b}) means that those terms $[z + i \nu_{q}]$ in Eq.~(\ref{eq5}) which specify the transient dynamics can be dropped.  Substituting Eq.~(\ref{eqA1}) into Eq.~(\ref{eq5}) yields

\begin{eqnarray} \label{eqA2}
&& \hat{K}_q(z)/(s^c_q z)= 1 + z \sum_k W^{(1)}_{q,k} \hat{g}_k(z)-D_q[P_\varepsilon] \nonumber\\
&& +z\sum_{kp} W^{(2)}_{q,kp} LT [g_k (t) g_p (t)](z) \nonumber\\
&& +[z \sum_k W^{(1)}_{q,k} \hat{g}_k (z)]^2 +\mathcal{O}_\varepsilon  \quad.
\end{eqnarray}

Using Eq.~(\ref{eqA2}) one obtains a corresponding expansion of $\hat{E}_q(z)=-\hat{K_q}(z)/z^2$ (cf. Eq.~(\ref{eq6b})). Then transformation into the time domain leads to

\begin{eqnarray} \label{A3}
&& E(t)/s^c_q = 1-\sum_k W^{(1)}_{q,k} g_k(t) - D_q[P_{\varepsilon}] \nonumber\\
&& - \sum_{kp} W ^{(2)}_{q,kp} g_k (t) g_p (t) \nonumber\\
&& - \sum_{kp} W ^{(1)}_{q,k}  W^{(1)}_{q,p} \ LT^{-1} [z \hat{g}_k (z) \hat{g}_p (z)] (t) + \mathcal{O}_{\varepsilon} \ . \quad
\end{eqnarray}

Substituting this expression into the formula $\phi_q(t)=\exp(-E_q(t))$ (cf. Eq.~(\ref{eq6b})) and expanding the exponential function, one arrives at

  \begin{eqnarray} \label{A4}
&& \phi_q(t)/f^c_q=1+ s^c_q \sum_k W^{(1)}_{q,k} g_k (t) + s^c_q D_q [P_\varepsilon] \nonumber\\
&& + s^c_q \sum_{kp} W^{(2)}_{q,kp} g_k (t) g_p (t) \nonumber\\
&& s^{c}_q \sum_{kp} W ^{(1)}_{q,k} W^{(1)}_{q,p} LT^{-1} [z \hat{g}_k (z) \hat{g}_p (z)](t) \nonumber\\
&&+ \frac{1}{2} (s^{c}_q)^2 \sum_{kp} W^{(1)}_{q,k} W^{(1)}_{q,p} g_k(t) g_p(t) +\mathcal{O}_{\varepsilon} \ .
\end{eqnarray}

Equation~(\ref{eq14a}) reads $\phi_q(t)/f^c_q-1=[s^c_q/f^c_q] g_p(t)$. As a result, one obtains the desired reformulation of the equations of motion

\begin{equation} \label{A5}
\sum_k [\delta_{q,k} - A^c_{q,k} ] g_k (t) = I_q (t) \quad,
\end{equation}
with $(A^c_{q,k})$ the stability matrix (Eq.~(\ref{eq10d})) at the critical point.

The $M$ components of the array ${\bf I}(t)$ are

\begin{eqnarray} \label{A6}
&& I_q(t) = f^c_q D_q [P_{\varepsilon}] + \sum_{kp} A^{c}_{q,kp} g_k (t) g_p(t)\nonumber\\
&& + \sum_{kp} N^{c}_{q,kp} LT^{-1} [z \hat{g}_k (z) \hat{g}_p (z) ] (t) + \mathcal{O}_{\varepsilon}
\end{eqnarray}

with

\begin{equation} \label{A7}
A^{c}_{q,kp}=f^c_q W^{(2)}_{q,kp} +\frac{1}{2} s^c_q f^c_q W^{(1)}_{q,k} W^{(1)}_{q,p} \quad,
\end{equation}

\begin{equation} \label{A8}
N^{c}_{q,kp} =f^c_q W^{(1)}_{q,k} W^{(1)}_{q,p} \quad ,
\end{equation}
where $W^{(1)}_{q,k}$ and $W^{(2)}_{q,kp}$ are given by Eqs.~(\ref{eq16b}) and~(\ref{eq16c}), respectively.
\end{appendix}

%


\newpage

\begin{thebibliography}{99}
\bibitem{1} U. Bengtzelius, W. G\"otze and A. Sj\"olander, J. Phys. C{\bf17}, 5915 (1984)

\bibitem{2} W. G\"otze ``Complex Dynamics of Glass-Forming Liquids - 
A Mode-Coupling Theory'' (Oxford University Press, Oxford, 2009)

\bibitem{3} D. Forster, ``Hydrodynamic Fluctuations, Broken Symmetry, and Correlation Functions'' (Benjamin, San Francisco, 1975)

\bibitem{4} J.P. Hansen and I.R. McDonald, ``Theory of Simple Liquids'' (Academic Press, Waltham, 2006)

\bibitem{5} G. Szamel, Phys. Rev. Lett. {\bf 90}, 228301 (2003)
\bibitem{6} J. Wu and J. Cao, Phys. Rev. Lett. {\bf95}, 078301 (2005)
\bibitem{7} P. Mayer, K. Miyazaki and D.R. Reichmann, Phys. Rev. Lett. {\bf 97}, 095702 (2006)
  \bibitem{8} K. Kawasaki and B. Kim, Phys. Rev. Lett. {\bf 86}, 3582 (2001)

  \bibitem{9} K.S. Schweizer and E.J. Saltzman, J. Chem. Phys. {\bf119}, 1181 (2003)

 \bibitem{10} G. Biroli and J.-P. Bouchaud, Europhys. Lett. {\bf 67}, 21 (2004)

 \bibitem{11} A. Ikeda and K. Miyazaki, Phys. Rev. Lett. {\bf104}, 255704 (2010), {\bf 106}, 049602 (2011)

 \bibitem{12} B. Schmid and R. Schilling, Phys. Rev. E{\bf81}, 041502 (2010); R. Schilling and B. Schmid, Phys. Rev. Lett. {\bf 106}, 049601 (2011)

   \bibitem{13} M. Tokuyama, Physica A{\bf395}, 31 (2014)

   \bibitem{14} M. Tokuyama and H. Mori, Prog. Theoret. Phys. {\bf55}, 411 (1976)

   \bibitem{15} Y. Kimura and M. Tokuyama, arXiv: cond-mat/1407.3921

 \bibitem{16} W. G\"otze and L. Sj\"ogren, J. Phys. C{\bf17}, 5759 (1984)

 \bibitem{17} W. G\"otze, E. Leutheusser and S. Yip, Phys. Rev. A{\bf23}, 2634 (1981); A{\bf 24},
 1008 (1981)

\bibitem{17a} S.K. Schnyder, F. H\"ofling, T. Franosch and Th.Voigtmann, J.Phys.: Condens. Matter {\bf23}, 234121 (2011)

\bibitem{18} W. G\"otze and R. Haussmann, Z. Phys. B{\bf72}, 403 (1988)

\bibitem{19} L. Fabbian, W. G\"otze, F. Sciortino, P. Tartaglia and F. Thiery, Phys. Rev. E{\bf59}, R1347 (1999); E{\bf 60}, 2430 (1999)

\bibitem{20} K. Dawson, G. Foffi, M. Fuchs, W. G\"otze, F. Sciortino, M. Sperl, P. Tartaglia, Th. Voigtmann and E. Zaccarelli, Phys. Rev. E{\bf63}, 011401 (2001)

\bibitem{21} J. Bergenholtz and M. Fuchs, Phys. Rev. E{\bf59}, 5706 (1999)
\end{thebibliography}

\begin{thebibliography}{73}%
\makeatletter
\providecommand \@ifxundefined [1]{%
 \@ifx{#1\undefined}
}%
\providecommand \@ifnum [1]{%
 \ifnum #1\expandafter \@firstoftwo
 \else \expandafter \@secondoftwo
 \fi
}%
\providecommand \@ifx [1]{%
 \ifx #1\expandafter \@firstoftwo
 \else \expandafter \@secondoftwo
 \fi
}%
\providecommand \natexlab [1]{#1}%
\providecommand \enquote  [1]{``#1''}%
\providecommand \bibnamefont  [1]{#1}%
\providecommand \bibfnamefont [1]{#1}%
\providecommand \citenamefont [1]{#1}%
\providecommand \href@noop [0]{\@secondoftwo}%
\providecommand \href [0]{\begingroup \@sanitize@url \@href}%
\providecommand \@href[1]{\@@startlink{#1}\@@href}%
\providecommand \@@href[1]{\endgroup#1\@@endlink}%
\providecommand \@sanitize@url [0]{\catcode `\\12\catcode `\$12\catcode
  `\&12\catcode `\#12\catcode `\^12\catcode `\_12\catcode `\%12\relax}%
\providecommand \@@startlink[1]{}%
\providecommand \@@endlink[0]{}%
\providecommand \url  [0]{\begingroup\@sanitize@url \@url }%
\providecommand \@url [1]{\endgroup\@href {#1}{\urlprefix }}%
\providecommand \urlprefix  [0]{URL }%
\providecommand \Eprint [0]{\href }%
\providecommand \doibase [0]{http://dx.doi.org/}%
\providecommand \selectlanguage [0]{\@gobble}%
\providecommand \bibinfo  [0]{\@secondoftwo}%
\providecommand \bibfield  [0]{\@secondoftwo}%
\providecommand \translation [1]{[#1]}%
\providecommand \BibitemOpen [0]{}%
\providecommand \bibitemStop [0]{}%
\providecommand \bibitemNoStop [0]{.\EOS\space}%
\providecommand \EOS [0]{\spacefactor3000\relax}%
\providecommand \BibitemShut  [1]{\csname bibitem#1\endcsname}%
\let\auto@bib@innerbib\@empty
\end{thebibliography}
\end{document}